\documentclass[fleqn,12pt,twoside]{article}
\usepackage{espcrc1}

\newcommand{\AmS}{{\protect\the\textfont2
  A\kern-.1667em\lower.5ex\hbox{M}\kern-.125emS}}

\title{ Cluster Properties and Particle Production in Relativistic 
Quantum Mechanics}

\author{W. N. Polyzou\address{Department of Physics and Astronomy\\ 
        The University of Iowa \\ 
        Iowa City, IA, 52246}%
\thanks{This work  was supported in part by 
the Department of Energy, Nuclear Physics Division, under contract DE-FG02-86ER40286.}}
       
\begin{document}

\maketitle

\begin{abstract}

I formulate a class of relativistic quantum mechanical models that
satisfy the cluster property and allow particle production.  The
models have a finite number of bare-particle degrees of freedom.  The
class of models include relativistic isobar models and the relativistic
Lee model.  I discuss elements of the construction that may be relevant
to treat the general case of an infinite number of degrees of freedom.

\end{abstract}
\bigskip

The Cluster property provides the justification for the few-body
programs at modern accelerators, such as TJNAF.  Accurate measurement
and precise calculations of few-body observables are used to obtain a
detailed understanding of the dynamics of systems with a manageable
number of degrees of freedom.  The cluster property constrains the
dynamics of many-body systems in terms of the dynamics of the few-body
subsystems.

Since the cluster property is trivially realized in non-relativistic
quantum mechanics, the cluster property is not a cause of concern for
low-energy reactions.  For physics on distance scales of less than a
Fermi, a relativistic formulation of the quantum theory is essential
and the formulation of the cluster property is non-trivial.  The
requirement that the dynamical unitary representation of the Poincar\'e
group asymptotically factors into tensor products of subsystem
representations puts non-linear constraints on the infinitesimal
generators which cannot be satisfied without many-body interactions.

The problem of formulating relativistic quantum mechanical models
satisfying the cluster property has been solved for $N$-particle
systems \cite{so1}\cite{fc1}\cite{wk1}\cite{wp1}.  
In these theories the required many-body interactions are
generated by the few-body interactions.  Models which allow 
particle production have new difficulties associated directly with 
particle production because these theories generally 
have a infinite number of degrees of freedom.

In this work artificial conservation laws are introduced that permit
particle production without requiring an infinite number of degrees of
freedom.  The conservation laws provide a mechanism to separate the
difficulties associated with an infinite number of degrees of freedom
from the difficulties that arise directly from particle production.
 
Given the conservation laws it is possible to formulate a large class
of relativistic quantum models with a bounded number of bare-particle
degrees of freedom \cite{wp2} which allow particle
production and satisfy the cluster property.  Models in this class
include the relativistic Lee model, which was previously treated in
\cite{fc1}\cite{fuda}, and relativistic isobar models.

The models treated in \cite{wp2} are characterized by having
meaningful few-body problems; few-body interactions are fixed by
comparing accurate few-degree-of-freedom calculations to precise
experimental measurements.  The few-body interactions do not change
when the subsystem is embedded in a many-degree-of-freedom system.  As
in the $N$-particle case, the dynamics of proper subsystems in the
many-degree-of-freedom problem puts strong dynamical constraints on
the resulting dynamics.  These constraints generate additional many-
body interactions from the few-body interactions.

I formulate dynamical relativistic models \cite{wp1} by first constructing a
unitary representation of the Poincar\'e group on the model Hilbert
space, which is the orthogonal direct sum of tensor products of
bare-particle Hilbert spaces.  I use Clebsch-Gordon coefficients of
the Poincar\'e group the decompose the tensor products of irreducible
representations into direct integrals of irreducible representations.
The dynamics is included by adding to the invariant mass interactions
that commute with the spin and the operators that label and transform
vectors in each irreducible subspace.  This construction, which I call
Wigner's form of dynamics$^1$ leads to a dynamical representation of
the Poincar\'e group.  Dirac's forms of dynamics are realized as
special cases associated with specific realizations of the
Clebsch-Gordan coefficients.

For more than two degrees of freedom a dynamics based on Wigner's
forms of dynamics fails to satisfy the cluster property, even for
models with short-ranged interactions.  In these theories the cluster
property can be restored by utilizing a generalization of the Mayer
cluster expansion based on a partial ordering of subsystems and a $C^*$
algebra of operators whose unitary elements
preserve binding energies and $S$-matrix elements.  This $C^*$
algebra contains unitary elements that map representations of the
Poincar\'e group satisfying the cluster property to representations
that violate the cluster property.  These operators, which can be
constructed explicitly, and generalized cluster expansion methods are
used to formulate a recursive construction of a  unitary representation
of the Poincar\'e group that clusters into tensor products of
subsystem representations.
 
The number of degrees of freedom in models with production is
controlled by introducing fictitious conserved ``charges''.  These
charges replace particle number in the $N$-particle case.  Clusters
are associated with partitions of charges, but different partitions
into subsystems are formulated on different subspaces of the Hilbert
space, because some bare particles have several ``charges'' which
cannot be separated.  This leads to a modification\cite{wp2} of the
formulation of the cluster property.  It also leads to a more complex
partial ordering on subsystems and requires extensions of the $C^*$
algebra of asymptotic constants.  The specific problems that arise in
the variable number of particle case are treated in ref. \cite{wp2},
This paper also contains a detailed treatment of a theory having bare
nucleon, pion, delta, and rho meson degrees of freedom.

There number of degrees of freedom is bounded by requiring that the
net change of any particle is positive.  Allowing zero or negative
charges removes the unphysical restrictions on the dynamics at
the cost of introducing an infinite number of degrees of freedom.

The resulting models necessarily have many-body interactions which are
generated from subsystem interactions.  These interactions ensure the
factorization of the unitary representation of the Poincar\'e group into
tensor products of subsystem representations when the subsystems are
asymptotically separated.  These interactions also generate
model-dependent many-body contributions to tensor and spinor
densities, such as the hadronic current operator.  Understanding the
role of the many-body contributions to the current is important if
lepton scattering is used as a precision probe of hadronic structure
on sub-nuclear scales.

Current research is directed at removing the ``positive charge''
restriction.  In order to have a meaningful few-body problem in these
theories I choose to describe the theory in terms of physical particle
degrees of freedom.  I replace the partial ordering on charge by a
partial ordering on center-of-momentum energy.  While a number of
technical problems need to be addressed in order define the dynamics
in the general case, the solution to the ``positive charge problem''
provides useful framework for studying the general problem.


\begin{thebibliography}{9}
\bibitem{so1}S. N. Sokolov, Dokl. Akad, Nauk. SSSR 233,(1977)575.
\bibitem{fc1}F. Coester, W. N. Polyzou, Phys. Rev. {\bf D}26, (1982)1348.
\bibitem{wk1}W. H. Klink and  W. N. Polyzou, Phys. Rev. {\bf C}54, (1996)1189. 
\bibitem{wp1}W. N. Polyzou, J. Math. Phys. 43, (2002)6024, nucl-th/0201013.
\bibitem{wp2}W. N. Polyzou, Phys. Rev. {\bf C}68, (2003)015202, nucl-th/0302023. 
\bibitem{fuda}M. Fuda, Phys. Rev. {\bf D},41(1990)534.
\end{thebibliography}
\end{document}